\documentclass[fleqn,10pt]{wlscirep}
\title{Gaussian-Schell analysis of the transverse spatial properties of high-harmonic beams}

\author[1*]{David T. Lloyd}
\author[1,2]{Kevin O'Keeffe}
\author[1]{Patrick N. Anderson}
\author[1]{Simon M. Hooker}
\affil[1]{Department of Physics, University of Oxford, Clarendon Laboratory, Parks Road, Oxford OX1 3PU, UK}
\affil[2]{Department of Physics, Swansea University, Singleton Park, Swansea, SA2 8PP, UK}
\affil[*]{Corresponding author: david.lloyd@physics.ox.ac.uk}

\begin{abstract}
High harmonic generation (HHG) is an established means of producing coherent, short wavelength, ultrafast pulses from a compact set-up. Table-top high-harmonic sources are increasingly being used to image physical and biological systems using emerging techniques such as coherent diffraction imaging and ptychography. These novel imaging methods require coherent illumination, and it is therefore important to both characterize the spatial coherence of high-harmonic beams and understand the processes which limit this property. Here we investigate the near- and far-field spatial properties of high-harmonic radiation generated in a gas cell. The variation with harmonic order of the intensity profile, wavefront curvature, and complex coherence factor is measured in the far-field by the SCIMITAR technique. Using the Gaussian-Schell model, the properties of the harmonic beam in the plane of generation are deduced. Our results show that the order-dependence of the harmonic spatial coherence is consistent with partial coherence induced by both variation of the intensity-dependent dipole phase as well as finite spatial coherence of the driving radiation. These findings are used to suggest ways in which the coherence of harmonic beams could be increased further, which would have direct benefits to imaging with high-harmonic radiation.\\ \\
This article was published in  \href{http://www.nature.com/articles/srep30504}{Scientific Reports}. \\Please cite as: Lloyd, D. T. et al. Gaussian-Schell analysis of the transverse spatial properties of high-harmonic beams. \emph{Sci. Rep.} 6, 30504; doi: 10.1038/srep30504 (2016).

\end{abstract}
\begin{document}

\flushbottom
\maketitle

\thispagestyle{empty}

\section*{Introduction}

The generation of high-order harmonics of a driving laser field via its nonlinear interaction with gaseous \cite{Brabec2000} or solid \cite{Vampa2015} media has been the subject of intense research since the early 1990s. A strong motivation for this work is the fact that high-harmonic generation (HHG) produces coherent radiation \cite{Bartels2002} in the extreme ultraviolet (XUV) and soft x-ray spectral region, where operation of conventional lasers is challenging \cite{SimonsBook}. The duration of HHG pulses has been shown to be as short as a few tens of attoseconds ($1\:\mathrm{attosecond}=1\times10^{-18}\:\mathrm{s}$) \cite{Goulielmakis2008,Hentschel2001}, well-matched to the natural time scale of atomic processes. Recently, the combination of high spatial and temporal coherence with short wavelength has allowed samples to be imaged using high harmonic beams at close to the Abbe limit, with a record resolution of 13.6 nm \cite{Tadesse}.

Characterization of the harmonic field serves two distinct purposes. On the one hand, quantification of the harmonic properties allows the physics of the laser-plasma interaction to be explored. For instance, strong-field processes like quantum phase interference \cite{Zair2008} can be encoded into the spatial properties of HHG. On the other hand, measuring the harmonics in space and time \cite{Kim2013} is crucial for applications requiring precise knowledge of the spatio-temporal structure of the field \cite{Calegari2016}. 

The spectral dependence of the spatial properties of HHG has been the subject of previous studies centred on specific components of the harmonic field. Ditmire \emph{et al.} measured the spatial coherence of high-order harmonics using a Young's slits arrangement \cite{Ditmire1996, Ditmire1997} and found that the dependence on harmonic order of the visibility of the fringe patterns was consistent with a small deviation from full coherence in the fundamental beam. While Le Deroff \emph{et al.} found in numerical simulations that the harmonic beam was only partially coherent, even in the case of a fully coherent driving beam and low levels of ionization; they attributed this behaviour to the spatial variation of the intensity-dependent phase of the harmonic dipole \cite{LeDeroff}. Frumker \emph{et al.} used the Spectral Wavefront Optical Reconstruction by Diffraction (SWORD) technique to characterize the wavefront and intensity profile of harmonics generated from molecular nitrogen \cite{Frumker2009,Frumker2012}. By assuming that the harmonics propagated as a Gaussian beam, the harmonic field in the plane of generation was deduced, showing that the source width decreased and wavefront curvature increased with increasing harmonic order. We note that Hartmann-Shack sensors have been used to measure the transverse coherence \cite{Schafer2002} and wavefront and transverse intensity profile \cite{Gautier2008} of high-harmonic beams, although this technique averages over the bandwidth of the incident radiation and, for the case of coherence measurements, requires a subsidiary measurement of the transverse beam profile.

Previous studies of the spatial properties of HHG have assumed that the radiation source is either fully coherent \cite{Frumker2012} or completely incoherent \cite{Ditmire1996}. We extend these treatments by interpreting our results within the more general Gaussian-Schell model (GSM) for the propagation of light from partially coherent sources \cite{Friberg1982}. Short wavelength radiation from both synchrotrons and free electron lasers has been analysed using the GSM \cite{Vartanyants2010}, however to our knowledge HHG sources have not been described using this approach.

In this paper we report the results of experiments using the SCIMITAR technique to measure the variation with harmonic order $q$ of the intensity width, wavefront curvature, and complex coherence factor (CCF) in the far-field. In particular, this approach allows us to investigate the physical processes which degrade the spatial coherence of the harmonic beam. We find that good agreement between the inferred source coherence width and an analytic model is achieved when the effects of both inherited partial coherence from the driver beam and spatio-temporal variation of the intensity-dependent phase of the induced harmonic dipole are included.

\section*{Methods}

\begin{figure}[ht]
\centering
\fbox{\includegraphics[width=13cm]{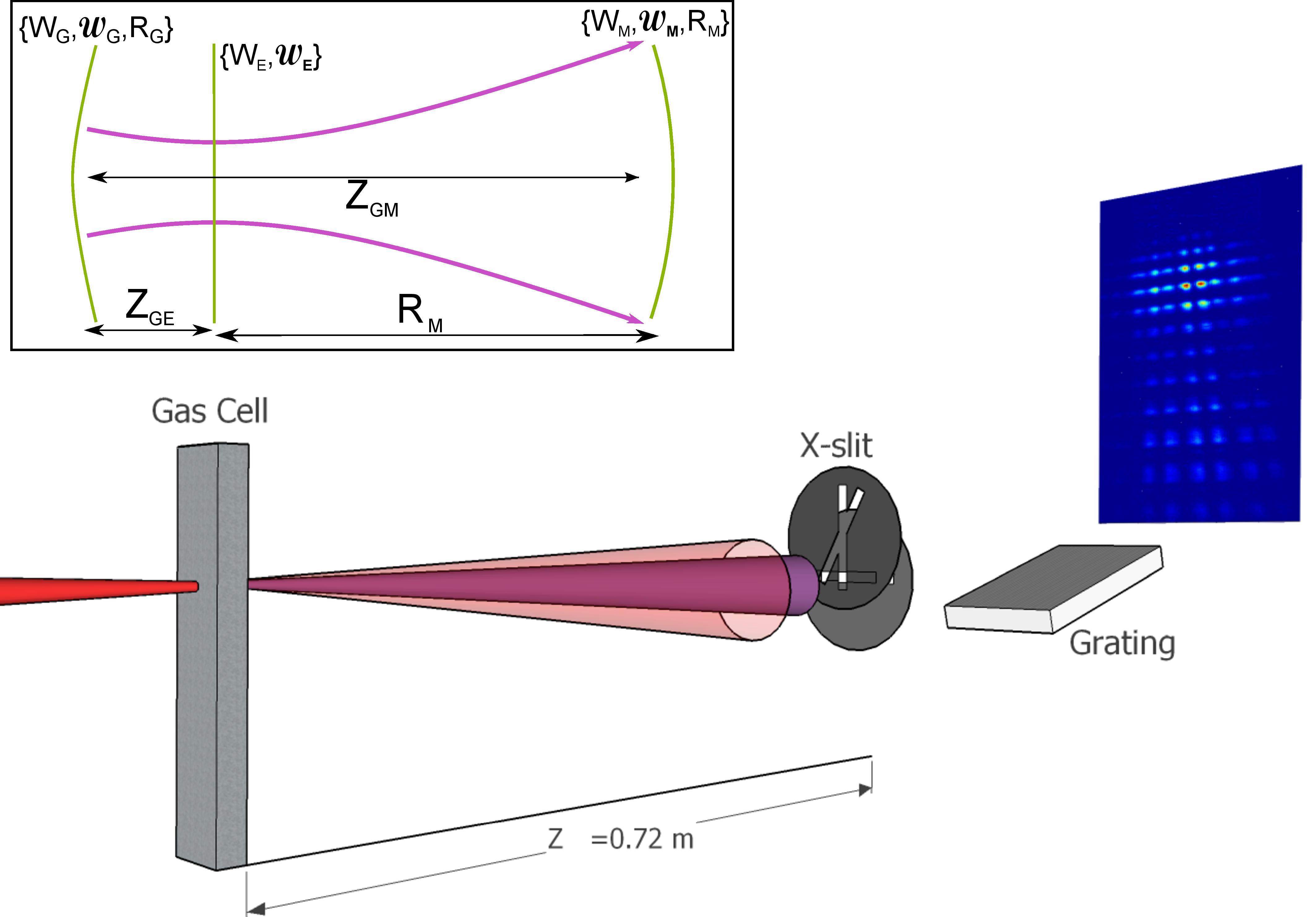}}
\caption{Schematic diagram of the experimental arrangement with relative separations exaggerated for clarity. The fundamental and harmonic beams are shown in red and purple respectively. The inset represents the evolution of the harmonic beam diameter (purple contour) and wavefront (green solid lines) with propagation distance. The notation employed is defined in the sketch and is as follows: subscript `G' for generation plane; subscript `E' for effective source plane; subscript `M' for measurement plane.}
\label{EXP}
\end{figure}

The SCIMITAR technique can be used to measure the spatial properties of a beam from a single scan. It has been described in detail elsewhere \cite{Lloyd2013} but, briefly, operates as follows: the spatial properties of the field are encoded into a series of interference patterns produced by a variable separation pinhole pair. Practically, the pinhole pair can be formed by the combination of a tilted `X' shaped slit placed in front of a horizontal slit. The horizontal pinhole separation can then be adjusted by moving the ‘X’ slit vertically; the tilt of the X-slit ensures one pinhole remains stationary throughout a scan. An imaging spectrometer is used to measure the resultant fringe patterns, and thus the spatial properties, as function of wavelength. We evaluate the fringe visibility at the centre portion of the resultant fringe pattern, thereby avoiding any reduction in the visibility caused by the finite temporal coherence of the harmonics \cite{Zuerch}. Further, since SCIMITAR records both the fringe visibility and transverse intensity profile in a single scan, it is possible to reconstruct the complex coherence factor $\mu(x_1,x_2)$. Measurement of the full complex coherence factor $\mu(x_1,x_2)$, where all possible combinations of $\{x_1,x_2\}$ within a given range are evaluated, is possible with SCIMITAR, but requires multiple measurements, each with a different static pinhole location. Alternate interferometric techniques \cite{Mang_thesis,Mang2014} are available for performing this sort of measurement more quickly. The experimental arrangement for SCIMITAR is depicted in figure \ref{EXP}.

For the experiment reported here, laser pulses from a Ti:sapphire laser system operating at 1 kHz with a centre wavelength of 800 nm and duration of approximately 40 fs, were spectrally broadened in a 1 m long argon-filled, differentially pumped, capillary and subsequently compressed to a duration of $\approx15$ fs by a set of chirped mirrors. No ionization was observed at the capillary entrance under operational conditions, the beam leaving the HCF  had good mode quality, and the compressed driving laser pulses yielded clean, unstructured FROG traces, with a small FROG error. The driving laser – and hence the harmonics it generated – is therefore expected to have been linearly polarized to a high degree. The pulses were directed through a 1 mm thick window into a vacuum chamber where the pulse energy was measured to be 180 $\mu$J. The beam was focussed by a spherical mirror with a focal length of 0.375 m. Astigmatism was minimised by ensuring that the incoming beam was at near-normal incidence to the focussing mirror ($1.6^{\circ}$ from the mirror normal). The focal spot diameter was measured to be $44\:\mu$m at low power and at atmospheric pressure. A gas cell comprising a thin-walled (0.1 mm thick) nickel tube pressed to an outer thickness of 1.4 mm, and with entrance and exit holes machined by the driving laser, was placed close to the laser focus and back-filled with argon at a pressure of 83 mbar. The generated harmonics subsequently propagated freely a distance 0.72~m to the SCIMITAR apparatus. Thin metallic filters were employed to prevent the driving radiation reaching the spectrometer: for the main experiment two 200-nm-thick Al filters were used allowing harmonic orders $q=23 - 43$ to be studied simultaneously. However runs in which the Al filters were replaced with a single 200-nm-thick Zr filter showed that up to $q=47$ was generated under identical experimental conditions.

\section*{Order Dependence of Harmonic Spatial Properties}

A SCIMITAR scan can determine three properties of the beam in the plane of the measurement:
the beam intensity width ($W_\mathrm{M}$),
the wavefront radius of curvature ($R_\mathrm{M}$) and
and the width of the complex coherence factor (or `coherence width') ($\mathcal{W}_\mathrm{M}$). We use the subscript `M' to indicate a quantity measured in the plane of the SCIMITAR pinholes. In our study (for a given harmonic order) the quantities $W_\mathrm{M}$ and $\mathcal{W}_\mathrm{M}$ correspond to full width at half maximum (FWHM) measures and are found by fitting Gaussian functions to the intensity profile and CCF, respectively. For $R_\mathrm{M}$, the recovered spatial phase profile was fitted to a function $\Phi(X)=\frac{k_qX^2}{2R_\mathrm{M}}$, where $k_q$ is the harmonic wavenumber and $X$ is the transverse distance from the beam axis in the measurement plane.

\subsection*{Intensity Width}

\begin{figure}[htbp]
\centering
\fbox{\includegraphics[width=17.5cm]{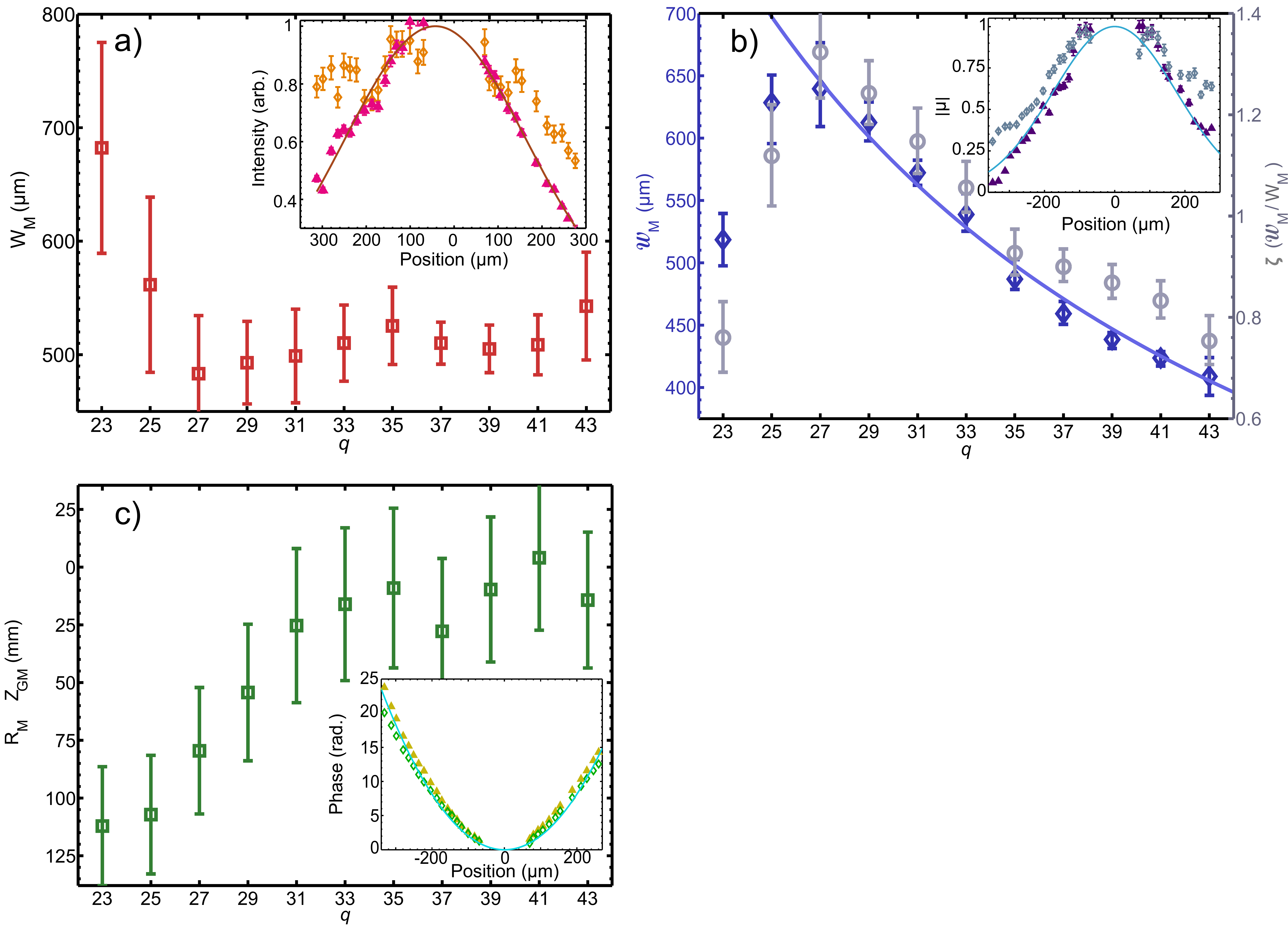}}
\caption{\textbf{a)} Variation of the measured intensity width $W_\mathrm{M}$ with $q$ (open red squares). Inset: comparison between the intensity profiles for $q$=23 (open orange diamonds) and $q$=41 (filled pink triangles). A Gaussian fit to $q$=41 is shown by the brown line. \textbf{b)} Variation of harmonic coherence width $\mathcal{W}_\mathrm{M}$ with $q$ (open blue diamonds). A fit to orders 27--43 of the 1/$q$ dependence predicted by the VCZ theorem is shown by the solid mauve line. The normalised coherence width $\zeta$ is plotted on the right hand axis with grey circles. Inset: CCF magnitude for $q$=41 (filled purple triangles) with a Gaussian fit (light blue solid line). The open grey diamonds show the CCF magnitude for $q$=23. \textbf{c)} The measured wavefront curvature $R_\mathrm{M} - Z_\mathrm{GM}$ is plotted as a function of $q$ (open green squares). Inset: Spatial phase profile of order $q$=41 (open dark green diamonds) with a fitted parabola (turquoise solid line). The same quantity for $q$=23 is shown with filled yellow triangles. The error bars are smaller than the symbol size for all data points.}
\label{FF}
\end{figure}

Figure \ref{FF} a) shows $W_\mathrm{M}$ as a function of $q$ for $q=23 - 43$. It is seen that two distinct regions may be identified: for $q \ge 27$ the width of the harmonic beam is almost independent of $q$; whereas for lower-order harmonics the width increases rapidly with decreasing $q$. The inset to figure \ref{FF} a) shows that the transverse profile of $q = 23$ is broad and asymmetric compared to the narrower, symmetric profile of $q = 41$, which is representative of the profiles measured for harmonics $q=27 - 43$. The larger scatter on the left side of the beam is reversed in the plot of the CCF magnitude for $q=41$ found in the inset of figure \ref{FF} b). This behaviour is observed for the other harmonic orders.

\subsection*{Spatial Coherence Width}
Figure \ref{FF} b) shows the variation of the coherence width $\mathcal{W}_M$ with harmonic order. The variation of the CCF magnitude with pinhole separation $d=x_2-x_1$ is shown for $q$ = 23 and $q$ = 41 in the inset of Fig \ref{FF} b). 

According to  the van Cittert-Zernike theorem the FWHM coherence width at a distance $Z_\mathrm{GM}$ from an \emph{incoherent} source of radiation shaped like a disc of radius $a$ is given by:
\begin{equation}
\mathcal{W}_\mathrm{M}\approx \frac{0.7 Z_\mathrm{GM} \lambda_q}{a}\propto \frac{1}{q}
\end{equation}
where $\lambda_q$ is the harmonic wavelength and $Z_\mathrm{GM}=0.72 $ m for our experimental arrangement. A clear 1/$q$ dependence of $\mathcal{W}_M$ is observed for $q=27-43$, as indicated by the mauve line in figure \ref{FF} b), but this dependence is not followed by harmonics $q = 23$ and 25. From the fit shown in figure \ref{FF} b) the observed coherence width is found to be equivalent to that produced by an incoherent disc of diameter $a=37.0\pm0.3\:\mu$m. As expected, this diameter is smaller than the measured spot size of the driving beam. Here the quantity $a$ represents the size of the equivalent incoherent source, discussed in prior studies \cite{Ditmire1996}. Although the incoherent source size can be used as a convenient comparative metric to quantify spatial coherence, in reality the harmonic source is partially coherent, as evidenced by the observed low beam divergence (approximately 1 mrad) \cite{Collett1978}. A physically more realistic model which incorporates this aspect is described below.

Figure \ref{FF} b) also shows the `normalised coherence width' $\zeta = \mathcal{W}_\mathrm{M}/W_\mathrm{M}$ as a function of $q$: the larger the value of $\zeta$, the closer the radiation is to being fully spatially coherent. In these experiments this parameter is largest for $q=27$ for reasons discussed later.

\subsection*{Wavefront Curvature}
Figure \ref{FF} (c) shows the variation of the quantity $R_\mathrm{M} - Z_\mathrm{GM}$ with harmonic order $q$. It can be seen that $R_\mathrm{M}$ increases with $q$ for $q \lesssim 33$ and becomes approximately constant for larger $q$. For harmonics $q<29$, $R_M$ is, within errors, smaller than  $Z_\mathrm{GM}$, indicating that the harmonics are generated with negatively curved wavefronts. For higher-order harmonics $R_\mathrm{M} \approx Z_\mathrm{GM}$, suggesting that $Z_\mathrm{GM}$ is much larger than the Rayleigh range of the harmonic source. A qualitatively similar trend was reported in the work of Frumker et al. \cite{Frumker2012}.

\section*{Simple Model of the Spatial Coherence of the HHG Source}

\label{SimplePic}

Here we outline a one-dimensional treatment of spatial coherence of a harmonic beam and establish our notation. The electric field of a beam of radiation may be described by the analytic signal \cite{BornandWolf}:
\begin{equation}
U(x_i,t)=U_i=A_0\epsilon_{q}(x_i)\chi_{q}(t)\exp[\mathrm{i}\phi(x_i,t)]
\label{Field1}
\end{equation}
where $A_0$ is the maximum value of $U_i$, and $\epsilon_{q}$ and $\chi_{q}$ are real envelope functions for the spatial and temporal parts of the field, respectively, which we have assumed are separable. The complex coherence factor evaluated at the locations $x_1$ and $x_2$, can be expressed as \cite{BornandWolf}:
\begin{equation}
\mu_{12}=\frac{\langle U_1U_2^*\rangle}{\sqrt{\langle|U_1|^2\rangle\langle|U_2|^2\rangle}}
\label{mugen}
\end{equation}
where, to avoid clutter, we have omitted the time dependence of the fields explicitly. The angled brackets in equation \ref{mugen} denote a time average. When the time average spans of the order of the pulse duration, $\mu_{12}$ is the CCF of a single pulse. If the time average length is much longer than the pulse duration, the CCF corresponds to that of the ensemble of pulses measured within that span. In the experiments described here, each acquisition represents the sum of $\approx 40,000$ harmonic pulses, thus the measured CCF is that of an ensemble rather than any single pulse.

Combining equations \ref{Field1} and \ref{mugen} we find:
\begin{align}
\mu_{12}=\frac{\langle\chi_{q}(t)^2\exp[\mathrm{i}(\phi_1-\phi_2)]\rangle}{\langle \chi_{q}(t)^2\rangle}
\label{Mu0}
\end{align}
where $\phi_i=\phi(x_i,t)$ is the temporal phase.

In deriving an expression for the harmonic CCF we will assume that the generation region is thin and hence we will neglect any longitudinal effects such as absorption and phasematching. Following the work of Sali\'eres \emph{et al.} \cite{Salieres1997}, the temporal phase of harmonic $q$ can be approximated by: 
\begin{equation}
\phi_{q}\approx q\phi_{0}+\phi_\mathrm{dq}
\label{phiq}
\end{equation}
where $\phi_0$ is the phase of the fundamental and $\phi_\mathrm{dq}$ is the dipole or intrinsic intensity-dependent phase \cite{Lewenstein1994}. The dipole phase may be written as: $\phi_\mathrm{dq}\approx-\alpha_{q}^j I_\mathrm{0}$, 
where $\alpha_{q}^j$ is a coefficient which depends on the harmonic order and the electron trajectory $j$ associated with the harmonic emission, and $I_\mathrm{0}$ is the intensity of the fundamental beam at the time and position harmonic $q$ is generated.

Assuming the phase difference ($\Delta \phi=\phi_1-\phi_2$) is small allows us to use the truncated Taylor expansion of equation \ref{Mu0}. Discarding higher order terms and substituting in equation \ref{phiq}, the harmonic CCF can then be approximated by:
\begin{align}
|\mu_q|\approx 1-\frac{1}{2}q^2\bigg[\frac{\langle\chi_q^2\Delta\phi_0^2\rangle}{\langle\chi_q^2\rangle}-\frac{\langle\chi_q^2\Delta\phi_0\rangle^2}{\langle\chi_q^2\rangle^2}\bigg] -\frac{1}{2}\alpha^2\bigg[\frac{\langle\chi_q^2\Delta I_0^2\rangle}{\langle\chi_q^2\rangle}-\frac{\langle\chi_q^2\Delta I_0 \rangle^2}{\langle\chi_q^2\rangle^2}\bigg] 
-q\alpha_{q}^j \bigg[ \frac{\langle \chi_q^2 \Delta\phi_0 \Delta I_0 \rangle}{\langle\chi_q^2\rangle} + \frac{\langle \chi_q^2 \Delta \phi_0 \rangle \langle \chi_q^2 \Delta I_0 \rangle}{\langle \chi_q^2 \rangle^2} \bigg]
\label{halfway}
\end{align}
where $\Delta\phi_0=\phi_0(x_1,t)-\phi_0(x_2,t)$ and $\Delta I_0=I_0(x_1,t)-I_0(x_2,t)$. Equation \ref{halfway} can be rewritten in a more compact format:
\begin{align}
|\mu_q|\approx 1-\frac{1}{2}\bigg[q^2V_q^\prime(\Delta \phi_0)+{\alpha_{q}^j}^2 V_q^\prime(\Delta I_0)+2q\alpha_{q}^j C_q^\prime(\Delta \phi_0,\Delta I_0)\bigg]
\label{VarCovar}
\end{align}
where $V_q^\prime(F)$ and $C_q^\prime(F,G)$ can be thought of as the variance and covariance functions, respectively, weighted by the harmonic temporal profile $\chi_q^2$. Full expressions for $V_q^\prime(F)$ and $C_q^\prime(F,G)$ are found in the supplementary materials.

Writing the intensity difference as: $\Delta I_0=I_{00}\chi_0^2[\epsilon_0(x_1)^2-\epsilon_0(x_2)^2]$, where $I_{00}$ is the on-axis, peak \emph{driver} intensity, and replacing the variance of $\Delta\phi_0$ with the fundamental CCF $\mu_0^\prime$ (see supplementary materials), the harmonic CCF becomes:
\begin{align}
|\mu_q|\approx 1-q^2(1-|\mu_0^\prime|)-\frac{1}{2}{\alpha_{q}^j}^2I_{00}^2[\epsilon_0(x_1)^2-\epsilon_0(x_2)^2]^2V_q^\prime(\chi_0^2)-q\alpha_{q}^j I_{00}[\epsilon_0(x_1)^2-\epsilon_0(x_2)^2]C_q^\prime(\Delta \phi_0,\chi_0^2).
\label{mufinal}
\end{align}
 
The final two terms in equation \ref{mufinal} vanish for spatially symmetric driving fields when $x_1=-x_2$. Hence, measurements of the spatial coherence which employ a symmetric geometry — such as those presented by Ditmire et al. \cite{Ditmire1997} — are insensitive to dipole phase effects for spatially symmetric driving beams, as noted in previous theoretical work by Sali\'eres \emph{et al.} \cite{Salieres1997} The restriction of symmetric sampling of the CCF is removed in the present work since one pinhole was fixed at the centre of the beam (i.e. $x_1=0$).

\section*{The Gaussian-Schell Model}
\label{GSMsec}
The principal assumption of the Gaussian-Schell model (GSM) \cite{Friberg1982,Vartanyants2010} is that the cross-spectral density $\mathcal{C}(x_1,x_2,\omega)$ can be expressed as:
\begin{equation}
\mathcal{C}(x_1,x_2,\omega)=\sqrt{S_\mathrm{G}(x_1,\omega)S_\mathrm{G}(x_2,\omega)}\nu_\mathrm{G}(x_2,x_1,\omega)
\end{equation}
with
\begin{equation}
S_\mathrm{G}(x_i,\omega)=A\exp\bigg[{\frac{-x_i^2}{\sigma_\mathrm{G}(\omega)^2}}\bigg]
\end{equation}
and
\begin{equation}
\nu_\mathrm{G}(x_2,x_1)=\exp\bigg[{\frac{-(x_2-x_1)^2}{\eta_\mathrm{G}(\omega)^2}}\bigg]
\end{equation}
where $S_\mathrm{G}$ is the spectral density with an amplitude of $A$ and $\nu_\mathrm{G}$ is the spectral degree of coherence (SDC). The corresponding source intensity and coherence widths (FWHM) are given by $W_\mathrm{G}=2\sqrt{\ln(2)}\sigma_\mathrm{G}$ and $\mathcal{W}_\mathrm{G}=2\sqrt{\ln(2)}\eta_\mathrm{G}$, respectively. Here the subscript `$G$' denotes a property evaluated at the plane where the radiation was generated, rather than in the measurement plane (downstream). It can be shown that the CCF ($\mu_\mathrm{G}$) and the SDC ($\nu_\mathrm{G}$) are equivalent for a narrow frequency interval $\Delta\omega\ll\omega$ (as the case for a single harmonic order) \cite{BornandWolf}.

After propagation a distance $Z$ to the measurement plane, the spectral intensity and SDC take the following form \cite{Friberg1982,Vartanyants2010}:
\begin{equation}
S_M(X_i)=A^\prime(z)\exp\bigg[\frac{-X_i^2}{\sigma_\mathrm{M}^2}\bigg]
\end{equation}
\begin{equation}
\nu_M(X_2-X_1)=\exp\bigg[\frac{-(X_2-X_1)^2}{\eta_\mathrm{M}^2}\bigg]
\end{equation}
with
\begin{equation}
\sigma_\mathrm{M}^2= \sigma_\mathrm{G}^2+\frac{(4+\zeta^2)Z^2}{k_q^2\eta_\mathrm{G}^2}
\label{WPI}
\end{equation}
\begin{equation}
\eta_\mathrm{M}^2=\eta_\mathrm{G}^2+\frac{(4+\zeta^2)Z^2}{k_q^2\sigma_\mathrm{G}^2}
\label{WCI}
\end{equation}
where $A^\prime(z)$ is the new spectral amplitude, $X_i$ ($i\in\{1,2\}$) denotes a point on a plane transverse to the beam propagation direction and $k_q=\frac{2\pi}{\lambda_q}$ is the angular wavenumber. It can be shown that for GSM beams the normalised coherence width $\zeta$ is a constant of propagation, in other words $\zeta=\eta_\mathrm{M}/\sigma_\mathrm{M}=\eta_\mathrm{G}/\sigma_\mathrm{G}$.

The GSM gives the properties of a beam originating from a plane in which the phase of the radiation is invariant with transverse position. As noted above, however, the wavefronts of the harmonics are not in general expected to be planar at the source. To generalise the GSM to sources with curved wavefronts we first invert equations \ref{WPI} and \ref{WCI} to find the beam properties in the plane where the wavefronts are planar, a distance $R_\mathrm{M}$ upstream of the measurement plane. Quantities in this effective source plane are denoted with the subscript `E'. This inversion yields:
\begin{equation}
\sigma_\mathrm{E}=\frac{1}{\sqrt{2}}\Bigg[\sigma_\mathrm{M}^2-\frac{(k_q^2\eta_\mathrm{M}^2\sigma_\mathrm{M}^4-4R_\mathrm{M}^2\eta_\mathrm{M}^2-16R_\mathrm{M}^2\sigma_\mathrm{M}^2)^\frac{1}{2}}{k_q\eta_\mathrm{M}}\Bigg]^\frac{1}{2}
\end{equation}
\begin{equation}
\eta_\mathrm{E}=\frac{1}{\sqrt{2}}\Bigg[\eta_\mathrm{M}^2-\frac{\eta_\mathrm{M}(k_q^2\eta_\mathrm{M}^2\sigma_\mathrm{M}^4-4R_\mathrm{M}^2\eta_\mathrm{M}^2-16R_\mathrm{M}^2\sigma_\mathrm{M}^2)^\frac{1}{2}}{k_q\sigma_\mathrm{M}^2}\Bigg]^\frac{1}{2}.
\end{equation}

The same procedure then gives the beam properties in the generation plane as:
\begin{equation}
\label{W0I}
\sigma_\mathrm{G}=\sqrt{\sigma_\mathrm{E}^2+\frac{(4+\zeta^2)(R_\mathrm{M}-Z_\mathrm{GM})^2}{k_q^2\eta_\mathrm{E}^2}}
\end{equation}
\begin{equation}
\eta_\mathrm{G}=\sqrt{\eta_\mathrm{E}^2+\frac{(4+\zeta^2)(R_\mathrm{M}-Z_\mathrm{GM})^2}{k_q^2\sigma_\mathrm{E}^2}}
\label{W0C}
\end{equation}

\section*{Interrogating the High Harmonic Source}
\label{Source}

\subsection*{Source Size}
\label{WIsec}
\begin{figure}[htbp]
\centering
\fbox{\includegraphics[width=17.5cm]{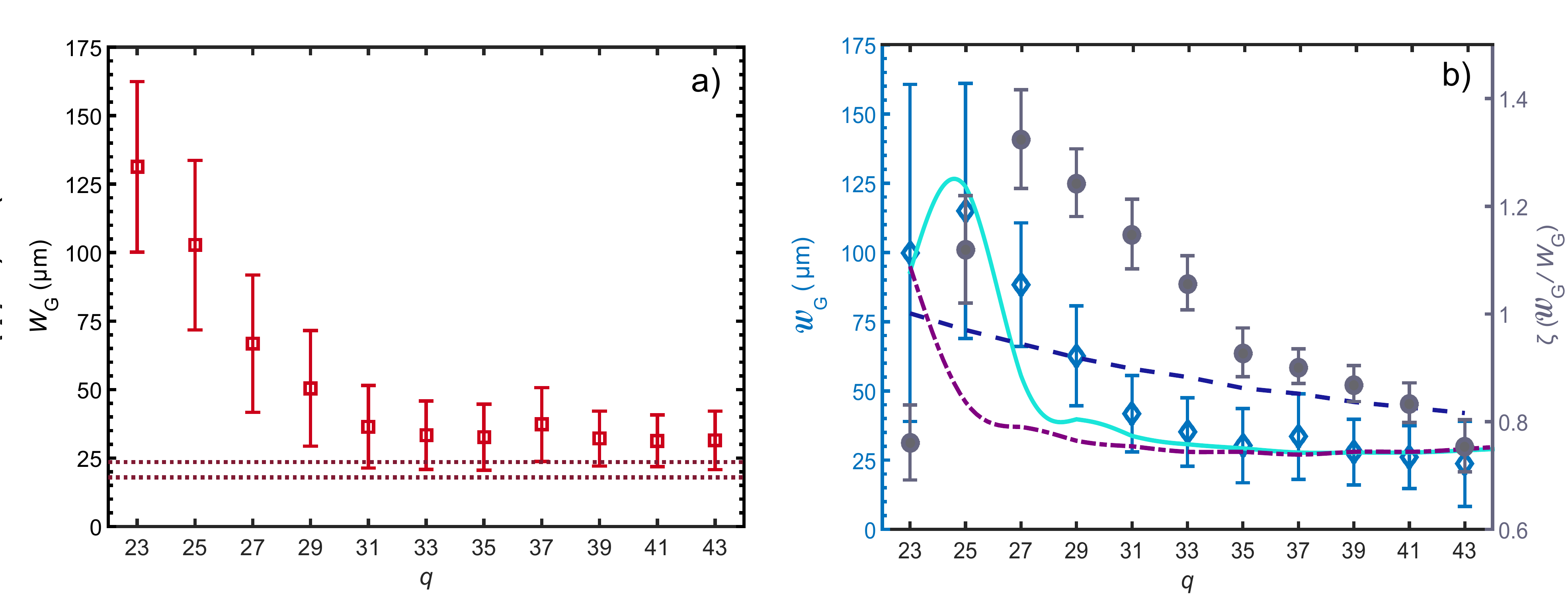}}
\caption{ \textbf{a)} Harmonic source intensity width $W_\mathrm{G}$ plotted as a function of order (open red squares). The brown dotted lines bound an interval consistent with equation \ref{sqrtn} and the SFA calculation. \textbf{b)} Harmonic source coherence width $\mathcal{W}_\mathrm{G}$ plotted as a function of order (open blue diamonds). The three lines corresponds to fits derived from equation \ref{mufinal} with $C^\prime(\Delta \phi_0,\chi_0)=0$ and: no dipole phase but a partially coherent driver (dashed navy blue line), dipole phase with a fully coherent driver (dot-dashed purple line) or the full expression — dipole phase + partially coherent driver (solid light blue line). The normalised coherence width ($\zeta$) is plotted with filled grey circles, these values are identical to those plotted in figure \ref{FF} by virtue of the properties of the Gaussian-Schell model.}
\label{NF}
\end{figure}

Figure \ref{NF} a) shows, as a function of $q$, the harmonic source intensity width $W_\mathrm{G}$ deduced from the GSM analysis. For both plots contained within figure \ref{NF} the error bars are calculated from propagation of the errors shown in figure \ref{FF}. In figure \ref{NF} a) it may be seen that for $q = 31 - 43$ the source width is approximately constant at $W_\mathrm{G}\approx 33\:\mu$m.

It has been shown previously \cite{Lewenstein1994} that the peak intensity of a harmonic order in the plateau region generated by a single atom can be approximated by: $I(q\omega_0)\propto I(\omega_0)^{n}$, with $n>1$ and $\omega_0$ refers to the angular frequency of the fundamental. Assuming a Gaussian transverse profile, within this model we expect:
\begin{equation}
W_\mathrm{G}=\frac{W_\mathrm{0}}{\sqrt{n}}
\label{sqrtn}
\end{equation}
where $W_\mathrm{0}$ is the fundamental intensity width (FWHM). Our calculations within the Strong Field Approximation (SFA) \cite{Lewenstein1994}, of a single argon atom driven by an intense 800 nm wavelength laser field, find values of $n$ in the region 3.5 -- 6. It is noted that if $I(q\omega_0)\propto I(\omega_0)^{n}$ holds, then the source size is independent of harmonic order, in so much as $n$ is -- as observed for $q=$31 -- 43.

The brown dotted lines in figure \ref{NF} a) show $\frac{W_0}{\sqrt{n}}$, with $W_0=44\:\mu$m and the upper and lower lines refer to $n=3.5$ and $n=6$, respectively. In spite of the simplicity of the model, agreement with the experimental values is reasonably good.

\subsection*{Source Coherence Width}
\label{WCsec}
Figure \ref{NF} b) shows, plotted as a function of $q$, $\mathcal{W}_\mathrm{G}$ deduced from the measured data and equation \ref{W0C}. Generally, $\mathcal{W}_\mathrm{G}$ decreases with increasing $q$. The data are fitted to equation \ref{mufinal} for three scenarios. For all fits it was assumed that the harmonic temporal profile $\chi_q^2$ was a top-hat function, however the width of $\chi_q^2$ varied with $q$ in the following way. For each harmonic the switch-on time $t_s(q)$ was taken to be the time at which the driving intensity was $\kappa$ times the threshold intensity $I_{0q}$ for generating the harmonic, which in turn was found from the cut-off law: $q \hbar\omega_0 = I_\mathrm{p} + C I_{0q}\lambda_0^2$, where $I_\mathrm{p}$ is the ionization potential of argon and $C$ is a constant. We use $\kappa$ as an order independent parameter which we fit to the data. We make the constraint $\kappa>1$ reflecting that a harmonic in the plateau is generated at a higher intensity, and hence at a later time on the leading edge of the pump pulse, than that dictated by the cut-off law. For all harmonic orders, generation was assumed to cease at $t_f = -2fs$, where the peak of the driver pulse occurs at $t = 0$, since an ADK calculation \cite{Ammosov1986} for the ionization in the medium shows that the on-axis ionization fraction is in excess of 0.3 at this time. Hence any emission for $t>t_\mathrm{f}$ is likely to be relatively weak owing to depletion and a rapidly decreasing coherence length. It should be noted that changes of the order of $\pm 1\:\mathrm{fs}$ to $t_\mathrm{f}$ had a negligibly small effect on the fitted curves. A summary of the three fit parameters is shown in table \ref{Fit summary}.

\begin{table}[ht]
\centering
\begin{tabular}{|l|l|l|l|}
\hline
Fit type & $\gamma$ & $\mathcal{W}_0$ ($\mathrm{\mu}$m) & $\kappa$ \\
\hline
Full Expression & $2.52\pm0.05$ & $3730\pm172$ & $1.19\pm0.02$  \\
No dipole phase & N/A & $2098\pm410$ & $1.00+0.01$\\
Dipole phase only & $2.20\pm0.03$ & N/A & $1.16\pm0.01$  \\
\hline
\end{tabular}
\caption{Summary of fit parameters corresponding to lines of best fit shown in figure \ref{NF}.}
\label{Fit summary}
\end{table}

When the role of the dipole phase is neglected (i.e. with $\alpha_{q}^j=0$), the finite harmonic coherence width arises from partial coherence in the fundamental alone. Assuming that the CCF of the driver is a Gaussian of FWHM $\mathcal{W}_0$, equation \ref{mufinal} gives:
\begin{align}
\label{Coh_only}
\mathcal{W}_\mathrm{G} = \mathcal{W}_0 \sqrt{\frac{\ln \left[\left(1 - 1/2q^2\right)^{-1} \right]}{\ln 2}}  \\
\approx \frac{\mathcal{W}_0}{q\sqrt{2\ln(2)}}
\end{align}
where the approximation holds in the limit $q\gg1$. As originally noted by Ditmire \emph{et al.} \cite{Ditmire1997}, only a very small departure from full coherence in the fundamental — corresponding to large values of $\mathcal{W}_0$ — is needed to produce a measurable reduction in the coherence of the harmonic field. A fit of equation \ref{Coh_only} is shown in figure \ref{NF} yields $\mathcal{W}_0=2.098\pm 0.410$ mm, which is much larger than the focal spot diameter. The fit also gives $\kappa=1.00^{+0.01}_{-0.00}$. We see that the agreement of this simple model with the data is poor.

Also shown in figure \ref{NF} b) are fits of two models in which the variation of $\phi_\mathrm{dq}$ is accounted for. In both cases $\alpha^j_q I_{00}$ is assumed to vary as $\alpha^j_q I_{00} = \beta + \gamma(q-q_\mathrm{cut-off})$, where $\gamma$ is a fit parameter and $q_\mathrm{cut-off} = 47$ is the order of the observed harmonic cut-off. The parameter $\beta$ was set equal to 57.2 so that, when combined with the estimated on-axis peak intensity of $I_{00}=4.4\times10^{14}\:\mathrm{W}\mathrm{cm}^{-2}$, the value of $\alpha^j_{47} $ was consistent with the previously reported value at the harmonic cut-off \cite{Auguste2009}.

The dot-dashed purple line shows a fit where the driver is assumed to be fully coherent and the harmonic partial coherence stems from variation of the dipole phase alone. Agreement is good for this fit at higher orders, with $\gamma = 2.20\pm 0.03$ and $\kappa=1.16\pm0.01$.

The solid light blue curve shows a fit in which the effects of the dipole phase and the finite coherence of the driver are both included yielding $\mathcal{W}_0=3.730 \pm 0.172$ mm, $\kappa=1.19\pm0.02$ and $\gamma = 2.52\pm 0.05$. For this fit the covariance term in equation \ref{mufinal} was neglected. It can be seen that the fit is in very good agreement with the data.

It is clear that the deduced variation of the coherence width in the generation plane is not consistent with the effects of either finite driver coherence or intensity-dependent dipole phase alone. However a simple model which includes both of these effects is able to reproduce the harmonic dependence of the spatial coherence width quite closely.

\section*{Discussion and Conclusions}
\label{Conc}
In summary we have measured the far-field intensity profile, wavefront curvature, and complex coherence factor magnitude for high-order harmonics generated by 15 fs duration, 800 nm wavelength pulses. We find that for orders $q\ge27$, $W_\mathrm{M}$ is roughly independent of $q$, while the $\mathcal{W}_\mathrm{M}$ closely follows a $1/q$ fit. Orders $q=23$ and $q=25$ were found to possess significantly different spatial properties compared to the other orders measured, with the intensity profile notably asymmetric. The origin of this effect is not known, but we note that in the case of orders $q=23$ and $q=25$ the absorption length in argon was smaller than the longitudinal length of the gas cell, which was not the case for the higher-order harmonics.

The properties of the harmonics in the generation plane were deduced from the measured quantities by applying a Gaussian-Schell analysis, which, to our knowledge, is the first time this approach has been used for high-harmonic radiation. We find that $W_\mathrm{G}$ initially decreases with $q$ before settling to a value in reasonable agreement with the predictions of strong-field theory.

It might be expected that $W_\mathrm{M}$ would decrease with increasing $q$, given the near constancy of the source size $W_\mathrm{G}$ and the decrease in the harmonic wavelength with $q$. Instead we measure $W_\mathrm{M}$ to be approximately constant for orders $q>27$. This unexpected behaviour stems from the fact that the coherence width in the source plane $\mathcal{W}_\mathrm{G}$ decreases with $q$ for $q>27$. The poorer coherence of the higher orders tends to increase the divergence of the harmonic, and hence the downstream beam size, and this effect approximately balances the effect of the decreasing wavelength. Non-symmetric sampling of the beam ensured that the measurements of $\mathcal{W}_\mathrm{G}$ were sensitive to the effects of dipole phase. We find that the partially coherent harmonic emission cannot be satisfactorily explained as being inherited from partial coherence in the driver alone. Rather, a simple model invoking both driver partial coherence and the spatio-temporal variation of the dipole phase yielded excellent agreement over the span of harmonic orders we measured.

We note that our treatment assumed a thin generation region. The confocal parameter of the driving radiation was approximately 11 mm, compared to a cell length of 1.2 mm; as such the transverse intensity profile of the driving radiation would have been nearly the same throughout the cell. We estimate that with our experimental parameters the coherence length ($L_\mathrm{c} = \pi/|\Delta k|$), was longer than the gas cell for $q < 39$, and comparable to the cell length for the higher-order harmonics investigated. These values, and the good agreement between our 1-D model and the data, allow us to conclude that treating the generation region as thin was a reasonable approximation in this case.

The key quantity for experiments which utilize the spatial coherence of the beam is the normalised coherence width $\zeta$. In figure \ref{FF} b), $\zeta$ was found to be largest when $q=27$. Since (for a GSM beam) $\zeta$ is a constant of propagation, the same values also hold for the harmonic source [as evidenced in figure 3 b)]. Hence maximizing $\zeta$ in the generation plane amounts to optimizing it in any other plane. In this work both $W_\mathrm{G}$ and $\mathcal{W}_\mathrm{G}$ decrease with $q$, but they do so at a different rate and hence $\zeta$ was maximized for an
intermediate plateau order, in our case $q = 27$. Our measurements show that $\zeta$ decreased rapidly with $q$, and was less than unity for the highest orders investigated. This unfavourable scaling of $\zeta$ with $q$ suggests that harmonics of a very high order could have comparatively poor transverse coherence, potentially making them unsuitable for applications such as holography \cite{Bartels2002} and coherent diffraction imaging \cite{Tadesse}.

Information on the spectral dependence of the harmonic spatial properties could be used to improve the convergence of phase retrieval algorithms for lens-less imaging applications, in particular those using multiple harmonic wavelengths simultaneously \cite{Witte}. Further, our results indicate that harmonics with high $\zeta$ (i.e. near-spatially coherent) could be generated by a coherent driver with a top-hat spatial profile, compared to the more usual case of a Gaussian driving beam. Methods for increasing the spatial coherence of harmonic field by this, or other, means are of importance for the growing number of techniques requiring excellent spatial coherence from high harmonic beams. 


\section*{Supplementary Material}

\subsection{Theory of High Harmonic Spatial Coherence}
We write the electric field of harmonic order $q$ as:
\begin{align}
E_q(x_i,t)=A_q\epsilon_q(x_i)\chi_q(t)\exp[\mathrm{i}\phi_q(x_i,t)]
\label{SDEq1}
\end{align} 
where $\epsilon_q$ and $\chi_q$ are real, positive functions corresponding to envelopes in space and time, respectively, of the electric field, the subscript $q$ links the quantity explicitly with harmonic order $q$ and $x_i$ refers to transverse position in a plane where the harmonic is generated. Here, by using separable functions for the spatial and temporal parts of the harmonic field envelope, we have implicitly assumed no space-time coupling (STC) is present in the harmonic amplitude. Furthermore we concern ourselves with a one-dimensional harmonic source: it extends in the x-direction only.

The complex coherence factor parameterises the spatial coherence of the field. It can be written as:
\begin{align}
\mu_{q}=\frac{\langle E_q(x_1,t)E_q(x_2,t)^*\rangle}{\sqrt{\langle|E_q(x_1,t)|^2\rangle\langle|E_q(x_2,t)|^2\rangle}}
\label{SDEq2}
\end{align}
where the angle brackets denote a time average.
Substituting Eq.\ref{SDEq1} into \ref{SDEq2}:
\begin{align}
\mu_q=\frac{A_q^2\epsilon_q(x_1)\epsilon_q(x_2)\langle\chi_q(t)^2\exp\{i[\phi_q(x_1,t)-\phi_q(x_2,t)]\}\rangle}{A_q^2\epsilon_q(x_1)\epsilon_q(x_2)\langle\chi_q(t)^2\rangle}\nonumber\\
=\frac{\langle\chi_q(t)^2\exp\{i[\phi_q(x_1,t)-\phi_q(x_2,t)]\}\rangle}{\langle\chi_q(t)^2\rangle}
\end{align}
hence
\begin{align}
|\mu_q|=\bigg|\frac{\langle\chi_q(t)^2\exp\{\mathrm{i}\Delta\phi_q\}\rangle}{\langle\chi_q(t)^2\rangle}\bigg|
\label{Gen}
\end{align}
where $\Delta\phi_q=\phi_q(x_1,t)-\phi_q(x_2,t)$.
\subsection{Taylor Expansion of Phase Difference}
If the difference between the phases $\phi_q(x_1)$ and $\phi_q(x_2)$ is sufficiently small, a Taylor expansion may be used:
\begin{align}
\exp(\mathrm{i}\Delta\phi_q)=1+\mathrm{i}\Delta\phi-\frac{\Delta\phi^2}{2}...
\end{align}
substituting into equation \ref{Gen} and expanding out: 
\begin{align}
|\mu_q|\approx\bigg|\frac{\langle\chi_q(t)^2\bigg(1+\mathrm{i}\Delta\phi_q-\frac{\Delta\phi^2}{2}\bigg)\rangle}{\langle\chi_q(t)^2\rangle}\bigg|
=\bigg[1+ \frac{\langle\chi_q(t)^2\Delta\phi_q\rangle^2}{\langle\chi_q(t)^2\rangle^2}+ \frac{\langle\chi_q(t)^2\Delta\phi_q^2\rangle^2}{4\langle\chi_q(t)^2\rangle^2}  -\frac{\langle\chi_q(t)^2\Delta\phi_q^2\rangle}{2\langle\chi_q(t)^2\rangle}-\frac{\langle\chi_q(t)^2\Delta\phi_q^2\rangle}{2\langle\chi_q(t)^2\rangle}\bigg]^{\frac{1}{2}}.
\end{align}
Disregarding the higher order term proportional to $\langle\chi_q(t)^2\Delta\phi_q^2\rangle^2$ and applying a binomial expansion of the square root yields:
\begin{align}
|\mu_q|\approx 1-\frac{1}{2}\bigg(\frac{\langle\chi_q(t)^2\Delta\phi_q^2\rangle}{\langle\chi_q(t)^2\rangle}-\frac{\langle\chi_q(t)^2\Delta\phi_q\rangle^2}{\langle\chi_q(t)^2\rangle^2}\bigg)
\label{SDMod}
\end{align}
where the second term in equation \ref{SDMod} resembles the variance of $\Delta\phi_q$ \emph{modulated (or windowed) by the temporal envelope of the harmonic pulse}.

\subsection{Full Expression for the Harmonic CCF}
The harmonic phase can be written as $\phi_q=q\phi_0(x,t)-\alpha_j^q I_0(x,t)$, where $\phi_0$ and $I_0$ are the phase and intensity of the fundamental beam, respectively and $\alpha_j^q$ is an order dependent parameter related to the trajectory associated with the harmonic emission. The phase difference $\Delta\phi_q$ is then:
\begin{align}
\Delta\phi_q=q[\phi_0(x_1,t)-\phi_0(x_2,t)]-\alpha_j^q[I_0(x_1,t)-I_0(x_2,t)]\nonumber\\
=q\Delta\phi_0-\alpha_j^q\Delta I_0
\end{align}
where $\Delta I_0=I_0(x_1,t)-I_0(x_2,t)$.
Substituting into equation \ref{SDMod} yields:
\begin{align}
|\mu_q|\approx1-\frac{1}{2}q^2\bigg[\frac{\langle\chi_q^2\Delta\phi_0^2\rangle}{\langle\chi_q^2\rangle}-\frac{\langle\chi_q^2\Delta\phi_0\rangle^2}{\langle\chi_q^2\rangle^2}\bigg] -\frac{1}{2}(\alpha_j^q)^2\bigg[\frac{\langle\chi_q^2\Delta I_0^2\rangle}{\langle\chi_q^2\rangle}-\frac{\langle\chi_q^2\Delta I_0 \rangle^2}{\langle\chi_q^2\rangle^2}\bigg] -\frac{\langle \chi_q^2q\Delta\phi_0\alpha_j^q\Delta I_0\rangle}{\langle\chi_q^2\rangle}  +\frac{\langle \chi_q^2q\Delta\phi_0\rangle\langle\chi_q^2 \alpha_j^q\Delta I_0\rangle}{\langle\chi_q^2\rangle^2} .
\label{SDlong}
\end{align}
Eq. \ref{SDlong} can be expressed in a more compact form as:
\begin{align}
|\mu_q|\approx 1-\frac{1}{2}\bigg[q^2V_q^\prime(\Delta \phi_0)+(\alpha_j^q)^2V_q^\prime(\Delta I_0)+2q\alpha_j^q C_q^\prime(\Delta \phi_0,\Delta I_0)\bigg]
\end{align}
where 
\begin{align}
V_q^\prime(F)=\frac{\langle\chi_q^2(F)^2\rangle}{\langle\chi_q^2\rangle}-\frac{\langle\chi_q^2F\rangle^2}{\langle\chi_q^2\rangle^2} \\
\mathrm{and} \nonumber\\
C_q^\prime(F,G)=\frac{\langle \chi_q^2F G\rangle}{\langle\chi_q^2\rangle}-\frac{\langle \chi_q^2F\rangle\langle\chi_q^2 G\rangle}{\langle\chi_q^2\rangle^2}
\end{align}
can be thought of as the variance and covariance functions, respectively, weighted by the harmonic temporal profile $\chi_q^2$.
If, over the duration of the harmonic emission, the variation in the driver phase difference ($\Delta \phi_0$) is `statistically independent' of the variation in the driver intensity difference ($\Delta I_0$), or if either $\Delta \phi_0$ or $\Delta I_0$ are independent of time or zero, then $C_q^\prime(\Delta \phi_0,\Delta I_0)=0$.

Eq \ref{SDMod} can be modified to yield an expression for the driver CCF evaluated \emph{during the emission of harmonic $q$}:
\begin{align}
|\mu_0^\prime|\approx 1-\frac{1}{2}\bigg(\frac{\langle\chi_q(t)^2\Delta\phi_0^2\rangle}{\langle\chi_q(t)^2\rangle}-\frac{\langle\chi_q(t)^2\Delta\phi_0\rangle^2}{\langle\chi_q(t)^2\rangle^2}\bigg).
\label{SDmu0}
\end{align}
This CCF likely differs to the true CCF of the driver (i.e. $|\mu_0|$) which is evaluated over the entire duration of the driver pulse. Using equation \ref{SDmu0}, we can write the harmonic CCF as:
\begin{align}
|\mu_q|\approx 1-q^2(1-|\mu_0^\prime|)-\frac{1}{2}(\alpha_j^q)^2V_q^\prime(\Delta I_0)-q\alpha_j^q C_q^\prime(\Delta \phi_0,\Delta I_0).
\end{align}
Writing the intensity difference as $\Delta I_0=I_{00}\chi_0^2[\epsilon(x_1)^2-\epsilon(x_2)^2]$, where $I_{00}$ is the peak \emph{driver} intensity, the harmonic CCF becomes:
\begin{align}
|\mu_q|\approx 1-q^2(1-|\mu_0^\prime|)-\frac{1}{2}\alpha^2 I_{00}^2[\epsilon(x_1)^2-\epsilon(x_2)^2]^2V_q^\prime(\chi_0^2)-q\alpha_j^q I_{00}[\epsilon(x_1)^2-\epsilon(x_2)^2]C_q^\prime(\Delta \phi_0,\chi_0^2)
\label{SDmufinal}
\end{align}

\section*{Acknowledgements}

This work was supported by EPSRC (grant numbers EP/G067694/1 and EP/L015137/1).\\
\noindent The authors would like to thank Ian A. Walmsley for helpful discussions.

\section*{Author contributions statement}

D.T.L, K.O'K. and S.M.H conceived the experiment, D.T.L and K.O'K constructed the experimental apparatus, D.T.L conducted the experiment, D.T.L analysed the results, P.N.A performed supporting simulations.  All authors commented on and reviewed the manuscript. 

\section*{Additional information}

The authors declare no competing financial interests.

\end{document}